\documentclass[%
 reprint,
 amsmath,amssymb,
 aps,
]{revtex4-2}

\usepackage{graphicx}
\usepackage[bb=boondox]{mathalfa}
\usepackage{dcolumn}
\usepackage{bm}
\usepackage{subcaption}
\usepackage{amsmath}
\usepackage{mathtools}
\usepackage{physics}
\usepackage{comment}
\usepackage[normalem]{ulem}
\usepackage{tikz-network}
\usepackage{ragged2e} 
\usepackage{comment}
\usepackage{todonotes}
\usetikzlibrary{decorations.pathmorphing}
\DeclareCaptionJustification{justified}{\justifying}

\usepackage{hyperref}
\hypersetup{
	colorlinks=true,
	linkcolor=blue,
	citecolor=blue,
	filecolor=magenta,
	urlcolor=cyan
}

\newcommand{\IBMZRL}{IBM Quantum, IBM Research Europe - Zurich, CH-8803 R\"{u}schlikon, Switzerland}

\begin{document}

\title{Optimizing two-dimensional isometric tensor networks with quantum computers}

\author{Sebastian Leontica}
\email{sebastian.leontica.22@ucl.ac.uk}
\affiliation{\IBMZRL}
\author{Alberto Baiardi}
\email{alberto.baiardi@ibm.com}
\affiliation{\IBMZRL}%
\author{Julian Schuhmacher}
\affiliation{\IBMZRL}%
\author{Francesco Tacchino}
\affiliation{\IBMZRL}%
\author{Ivano Tavernelli}
\affiliation{\IBMZRL}%

\date{\today}
\UseRawInputEncoding

\begin{abstract}
We propose a hybrid quantum-classical algorithm for approximating the ground state of two-dimensional quantum systems using an isometric tensor network ansatz, which maps naturally to quantum circuits. Inspired by the density matrix renormalization group, we optimize tensors sequentially by diagonalizing a series of effective Hamiltonians. These are constructed using a tomography-inspired method on a qubit subset whose size depends only on the bond dimension. Our approach leverages quantum computers to enable accurate solutions without relying on approximate contractions, circumventing the exponential complexity faced by classical techniques. We demonstrate our method on the two-dimensional (2D) transverse-field Ising model, achieving ground-state optimization on up to 25 qubits with modest quantum overhead---significantly less than standard solutions based on variational quantum eigensolvers. Overall, our results offer a path towards scalable variational quantum algorithms in both noisy and fault-tolerant regimes.
\end{abstract}

\maketitle

\textit{Introduction}---Accurately simulating physical properties of strongly-interacting quantum many-body systems is a central challenge in computational physics.
A key advancement in this quest was the density matrix renormalization group (DMRG) for approximately solving the ground state problem of gapped Hamiltonians in one dimension~\cite{White1992, White1993, Schollwock2005, Schollwock2011}.
The algorithm searches for the optimal matrix product state (MPS), a variational class based on tensor networks, which approximates the set of area-law quantum states~\cite{Verstraete2008, Cirac2009, Orus2014}.
This breakthrough sparked broad numerical~\cite{Stoudenmire2010, Haegeman2011, Haegeman2012, Zauner2018, Rakovszky2022} and theoretical advancements~\cite{Schuch2011, Turner2011, Cecile2024, Leontica2025}, and led to a deep and comprehensive understanding of one-dimensional quantum systems.

Extending the success of tensor networks to higher dimensions and arbitrary geometries is an ongoing challenge.
A special class of tensor networks, the projected entangled-pair states (PEPS), provides a natural generalization of MPS~\cite{Verstraete2004, Jordan2008} to multi-dimensional quantum systems.
The key difficulty with PEPS is that, unlike MPS, they are not efficiently contractible~\cite{Schuch2007, Haferkamp2020}.
This leads to exponential runtime for computing expectation values of local observables exactly, a crucial subroutine for optimizing the ansatz and extracting relevant physical properties.
The isometric tensor network state (isoTNS) ansatz introduced in~\cite{Zaletel2020} attempts to bypass these difficulties by restricting the constituent tensors to isometries, mimicking the canonical form of MPS.
This constraint brings several advantages, such as efficient computation of the state norm, less redundancy, and a natural interpretation of the tensor network as a quantum circuit.
Proposed generalizations of DMRG~\cite{Lin2022, Hyatt2020} to isoTNS yield accurate numerical results for medium-sized problems.
Several works extend this ansatz to cover thermal states~\cite{Kadow2023}, fermionic statistics~\cite{Dai2024}, infinite system sizes~\cite{Wu2023} and exotic entanglement patterns~\cite{Liu2024, Soejima2020}.
Despite being easier to manipulate compared to generic PEPS, optimization schemes for isoTNS are inherently approximate and may give rise to uncontrolled errors in the worst case, or incur a cost for performing an exact contraction that grows exponentially in the system size.
A recent study confirms that the problem of computing exact expectation values of local observables in an isoTNS is BQP-complete~\cite{Malz2025}.
Efficient contractions are possible using dual-isometric tensors~\cite{Yu2024}, but it is not yet clear how their expressivity compares to that of unconstrained PEPS.

In this work, we design a hybrid quantum-classical algorithm for optimizing the isoTNS efficiently, with a computational cost that grows polynomially with both the bond dimension and the number of sites.
Through sequential optimizations of the tensors, the algorithm naturally inherits DMRG's hallmark rapid convergence and robustness.
We replace the approximate classical contractions required to calculate local observables with an efficient, local quantum state tomography-like protocol, leveraging insights from recent hybrid tensor-network literature~\cite{Schuhmacher2025}.
Noiseless simulations show that our approach can optimize isoTNS to any desired accuracy efficiently.
This efficiency guarantees scalability to large system sizes, bypassing known computational complexity limitations of fully classical algorithms.
Moreover, the circuit structure and the measurement overhead required to solve problems close to the so-called quantum utility scale~\cite{kim2023evidence} are within the capabilities of current quantum processors and substantially lower than those of approaches based on variational quantum eigensolvers (VQEs).
Hence, our proposed approach paves the way for a new route to achieve quantum advantage on ground-state problems of strongly correlated lattice Hamiltonians on near-term and early fault-tolerant devices~\cite{Lanes2025_QuantumAdvantage}.

\begin{figure}[t]
  \centering
  \includegraphics[width=0.48\textwidth]{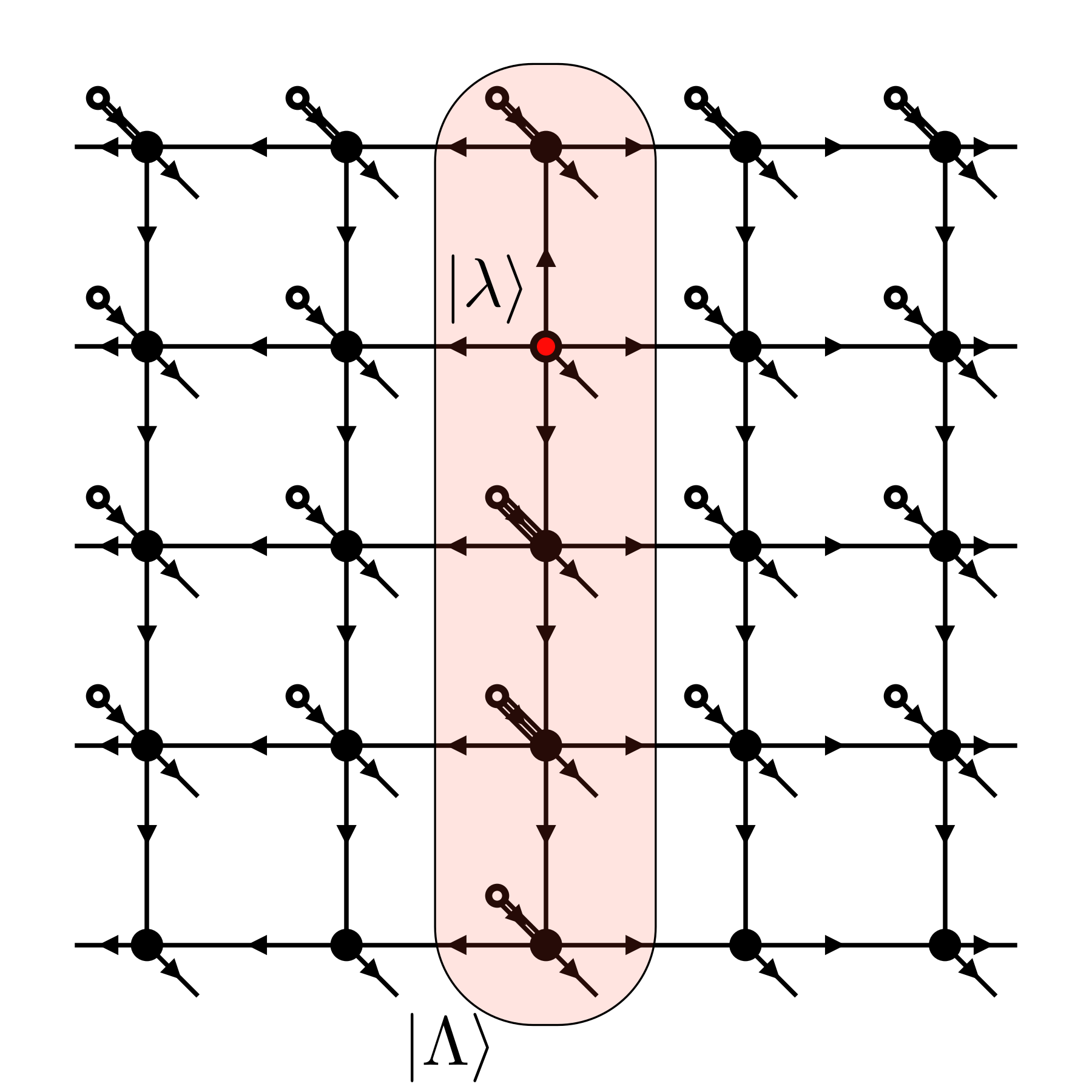} 
  \caption{Illustration of a $5\times5$ isoTNS.
  Black nodes denote unitaries with incoming (outgoing) arrows for inputs (outputs).
  The isometric center $\ket{\lambda}$ is drawn as a red node, and a transparent red overlay marks the corresponding central column $\ket{\Lambda}$.
  The white nodes denote indices fixed to some known reference value (corresponding to the $\ket{0}$ state in the quantum circuit).
  Uncontracted output legs are associated with physical qubits.
  Hence, this isoTNS encodes the state of a $7\times 5$ array of spins.}
  \label{fig:isoTNS}
\end{figure}

\textit{Methods}---An isometric tensor network is a restriction of more general PEPS.
It is formed by placing isometric tensors on the nodes of a grid and contracting neighboring tensors in a way that induces a well-defined causal structure.
This means that we can associate a direction with each edge such that the input-output structure of the nodes is respected and no loops are formed.
Every tensor has at least one free leg that indexes the state of a physical spin-$1/2$ particle located at the corresponding site.
Here, we will study 2D isoTNS placed on a square lattice, as illustrated in Fig.~\ref{fig:isoTNS}, although our techniques can be generalized to other topologies.
We refer to the root node in the causal structure as \emph{isometric center} and its corresponding column as \emph{central column}.
By contracting the tensor network along the directions indicated by the arrows, the information flows away from the isometric center both horizontally and along the central column.
Moreover, it flows downward along the remaining columns.
The presence of a casual order facilitates the mapping of the tensor network onto the quantum circuit, as discussed below.
The dimension of the virtual indices connecting neighboring nodes is called the bond dimension and will be denoted $D$.
The tensors located at the left and right edges of the lattice carry two physical legs (see Fig.~\ref{fig:isoTNS}).
This choice ensures that their output dimension is at least as large as the input dimension, a condition that must be met by any isometry.
With this convention, a $L_x \times L_y$ grid of tensors represents the state of a $(L_x+2)\times L_y$ array of spins.

Contracting the isoTNS along each column yields an MPS of local dimension $2^{L_y}$ and bond dimension $D^{L_y}$.
The isometry condition implies that this MPS is in canonical form~\cite{Schollwock2011}, with the central column $\Lambda$ representing the center of orthogonality.
Following~\cite{Zaletel2020}, we denote the columns to the left of the center as $A^{[x]}$ and the columns to the right as $B^{[x]}$, with $x$ the column index.
The isoTNS wave function $\Psi$ can be expressed in terms of MPS contractions as
\begin{equation}
    \Psi = A^{[1]} A^{[2]}\ldots A^{[c-1]}\Lambda^{[c]} B^{[c+1]} \ldots B^{[L_x]},
    \label{eq:isoTNS_as_MPS}
\end{equation}
where $c$ is the index of the central column.
From the isometric conditions of $A$ and $B$, it follows that the norm of $\Psi$ is given by the norm of the central column $\Lambda$.
In turn, this contracts further to the norm of the isometric center tensor $\lambda$.
This property will be crucial later in generalizing DMRG to isoTNS because it ensures that the optimization of the isometric center is well conditioned.
To this end, we denote the isometric map from the legs of the isometric center to the physical indices by $\mathcal{U}$ (obtained by contracting all black nodes in Fig.~\ref{fig:isoTNS}).
The vectorized isometric center $\ket{\lambda}$ is mapped to the target physical state through $\ket{\Psi} = \mathcal{U}\ket{\lambda}$.
The energy of the system in state $\Psi$ defined in Eq.~\eqref{eq:isoTNS_as_MPS} is given by
\begin{equation}
\label{eq:energy}
    E(\Psi) = \frac{\bra{\lambda} \mathcal{U}^\dagger H\mathcal{U}\ket{\lambda}}{\bra{\lambda}\ket{\lambda}} = \frac{\bra{\lambda}H_\text{eff}\ket{\lambda}}{\bra{\lambda}\ket{\lambda}},
\end{equation}
where $H_\text{eff}$ is an effective Hamiltonian acting only on the legs of the isometric tensor (i.e., $2 \times D^4$ dimensional for tensors located in the bulk of the network).
The vector $\ket{\lambda}$ minimizing the energy is the ground state of the effective Hamiltonian.
When combined with an algorithm for shifting the isometric center through the isoTNS, the local optimization of the energy~\eqref{eq:energy} yields a generalization of the DMRG to 2D systems \cite{Lin2022, Hyatt2020}.
This scheme is expected to retain key advantages of its one-dimensional counterpart \cite{Schollwock2005}, such as exponentially fast convergence and robustness against the presence of inaccurate local minima.

The primary challenge in scaling this algorithm to large problems lies in the computational cost of evaluating $H_\text{eff}$.
In general, this requires contracting a generic PEPS, which is typically performed via a boundary MPS contraction \cite{Lubasch2014}.
This approach is heuristic and may introduce uncontrolled errors, particularly for systems with large correlation lengths.
In the following, we design a quantum algorithm for constructing an unbiased estimator of the effective Hamiltonian efficiently.
The errors remain controlled even for systems near criticality, as we discuss in the Supplementary Material.
We also introduce a variant of this scheme -- inspired by the classical Lanczos method -- that estimates on the quantum processor the result of applying the Hamiltonian to an arbitrary vector, $H_\text{eff}\ket{v}$.
This can improve overall shot efficiency in some cases.

Our optimization protocol consists of 3 subroutines: shifting the isometric center through the network, constructing the effective Hamiltonian (or, alternatively, its action $H_\text{eff}\ket{v}$ on a given vector $\ket{v}$) and solving its eigenvalue problem.
The second step is performed on a quantum processor, with the remaining two executed on classical hardware.

\textit{Shifting the isometry center}---Two types of transformation are needed to shift the isometric center to a given position while maintaining the causal structure requirements outlined above.
The first one shifts the center along the central column (red shaded in Fig.~\ref{fig:isoTNS}), while the second one shifts the entire central column horizontally through the network.
The first transformation is exact and consists of a singular-value decomposition of the isometric center.
This is identical to shifting the center in an MPS \cite{Orus2014}.
The second transformation can only be performed approximately, as it is not known whether isoTNS with different central columns span the same variational space.
Fortunately, several heuristic methods have been proposed in the literature to efficiently implement this transformation with very high fidelity \cite{Lin2022, Zaletel2020}.

We shift the position of the central column horizontally using the sequential Moses move, introduced and investigated in recent works \cite{Lin2022, Zaletel2020}.
Consider, for instance, the shift of the central column to the right.
The Moses move approximately decomposes the central column as
\begin{equation}
    \Lambda^{[c]} \approx A^{[c]} \Lambda,
    \label{eq:MosesMove_1}
\end{equation}
where $A^{[c]}$ is a new isometric column (see Eq.~(\ref{eq:isoTNS_as_MPS})) and $\Lambda$ a second column of tensors without physical legs.
The new central column is found by contracting $\Lambda$ into the column at position $c+1$ to give
\begin{equation}
    \Lambda^{[c+1]} \approx \Lambda B^{[c+1]}.
    \label{eq:MosesMove_2}
\end{equation}

The overall error can be quantified using the fidelity between the starting columns $\Lambda^{[c]} B^{[c+1]}$ and the final columns $A^{[c]} \Lambda^{[c+1]}$ after the transformation.
Their overlap is a contraction between MPSs with bond dimension $D^2$, which can be evaluated efficiently.

\textit{Effective Hamiltonian}---We now describe the two methods for constructing and diagonalizing the effective Hamiltonian: the \emph{tomography method} for evaluating the entire effective Hamiltonian and the \emph{Lanczos method} for computing its action on a target tensor.
Both algorithms require sampling from a state that can be efficiently produced on a quantum computer.
As anticipated above, the latter strategy is more shot-efficient than the first one when implemented on quantum processors.

The \emph{tomography method} uses quantum state tomography to produce an estimator for the entire effective Hamiltonian.
This is done by first initializing a Bell pair for each leg of the isometric center (one for the physical leg and $\log_2 D$ for each virtual leg).
We then construct a quantum circuit realizing the isometry $\mathcal{U}$ obtained by removing from the isoTNS the isometric center tensor $\lambda$ (see Supplementary Material).
We apply the resulting circuit to the register composed of one qubit per Bell pair.
This yields a state over the joint ancilla-physical space given by
\begin{equation}
\label{eq:Psi}
    \ket{\Tilde{\Psi}} = \frac{1}{\sqrt{2D^4}}\sum_{i=0}^{2D^4-1}\ket{i}\otimes \mathcal{U}\ket{i},
\end{equation}
where $2D^4$ is the space dimension when the isometric center lies in the bulk and, hence, has one physical leg and four virtual legs.

Since Pauli operators $P$ form a complete basis for the operator space of the isometric center, the effective Hamiltonian can be decomposed as
\begin{equation}
    H_\text{eff} = \frac{1}{2D^4} \sum_P \Tr(H_\text{eff} P) P.
\end{equation}
If we construct $\Tilde\Psi$ and measure $P^T \otimes H$ on the resulting state, the resulting expectation value is a direct estimator of the required coefficients
\begin{equation}
    \langle P^T\otimes H \rangle_{\Tilde\Psi} = \frac{1}{2D^4}\Tr(H_\text{eff} P),
\end{equation}
where $H_\text{eff} = \mathcal{U}^\dagger H \mathcal{U}$ is the effective Hamiltonian we wish to estimate.

The coefficients are estimated by grouping the terms entering the Pauli decomposition of $H$ into sets of mutually commuting operators, preparing $\Tilde{\Psi}$, and measuring in the mutual eigenbasis of each set.
A detailed analysis of the sampling overhead required to estimate the effective Hamiltonian with a given accuracy is presented in the Supplementary Material.

The optimal isometric center $\ket{\lambda}$ is calculated by classical diagonalization of $H_\text{eff}$.
This step scales as $\mathcal{O}(D^{12})$ and is not the bottleneck in practice, as it is subleading compared to the tomography cost.

The Lanczos method reduces the shot cost of constructing $H_\text{eff}$ by directly targeting its lowest eigenstate.
This algorithm does not construct the full matrix $H_\text{eff}$, but only an oracle that applies $H_\text{eff}$ to arbitrary input vectors.
We propose using an iterative scheme, the Lanczos algorithm (notably, iterative schemes are routinely used also in classical DMRG-based methods).
With such an oracle, it is possible to construct the Krylov basis vectors defined recursively as $\ket{\lambda_n} = H_\text{eff}\ket{\lambda_{n-1}}$, starting from an initial guess $\ket{\lambda_0}$, and to find an approximate ground state of $H_\text{eff}$ in the reduced subspace.
The key advantage is the reduction in the number of estimated coefficients to only $\mathcal{O}(D^4)$ per new subspace vector, instead of all $\mathcal{O}(D^8)$ entries in the effective Hamiltonian.
Good approximations of the ground state can often be obtained with very small subspaces.

We compute $\ket{v'} = H_\text{eff}\ket{v}$ for arbitrary normalized $\ket{v}$ using a variant of the parameter-shift rule.
Let $V$ be a unitary operator that initializes $\ket{v}$ on a quantum register $V\ket{0} = \ket{v}$ (an optimal transpilation protocol is described in Ref.~\cite{Iten2016}).
If $\ket{s}$ denotes the space of all bitstrings defined on the same register, the vectors $V\ket{s}$ form a complete orthonormal basis.
We define the following set of input vectors for arbitrary integer $m$
\begin{equation}
    \ket{\phi_s^{m}} = \frac{1}{\sqrt{2}}\mathcal{U} V\left(\ket{0}+e^{i\frac{\pi}{2}m}\ket{s}\right).
\end{equation}
A simple calculation shows that the overlap of $H_\text{eff}\ket{v}$ onto the basis vector $V\ket{s}$ can be estimated as
\begin{equation}
  \bra{s}V^\dagger H_\text{eff}\ket{v} =
    \frac{1}{2} \sum_{m=0}^3 e^{i\frac{\pi}{2}m}\bra{\phi_s^m} H\ket{\phi_s^m},
  \label{eq:PS_Rule}
\end{equation}
when $s \neq 0$. For $s=0$ the overlap $\bra{v}H_\text{eff}\ket{v}$ is an expectation value and can be estimated directly.
Through Eq.~(\ref{eq:PS_Rule}), we can calculate the decomposition of $H_\text{eff} \ket{v}$ in the $V \ket{s}$ basis and recover the computational basis coefficients via the inverse operator $V^\dagger$.
As we detail in the Supplementary Material, the total number of samples required to obtain a relative energy error $\epsilon_r$ is $n_{tot} = \mathcal{O}(\xi^{2-\eta}D^8/\epsilon_r)$, where $\xi$ is the correlation length and $\eta$ the critical exponent characterizing the decay of 2-point correlators.

\begin{figure}[t]
  \centering
  \includegraphics[width=0.48\textwidth]{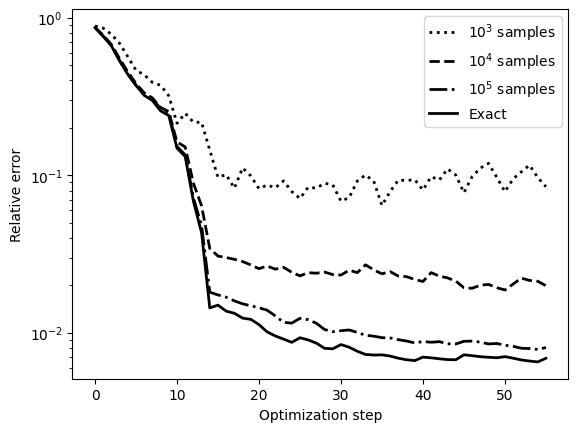}
  \caption{Result of optimizing a $5\times 5$, $g=3.5$ TFIM using the tomography method. We compare the optimization profile obtained using different numbers of shots per circuit to the result from exact tensor contraction. The number of samples is reported per tomography circuit executed during the effective Hamiltonian estimation phase. One sweep is composed of 15 optimization steps. The relative error is calculated with respect to the exact diagonalization result.}
  \label{fig:resultsTomo}
\end{figure}

\textit{Results}---We benchmark the algorithm on the 2D transverse-field Ising model $H = -\sum_{\langle i,j\rangle} \sigma_i^z \sigma_j^z -g\sum_i \sigma_i^x$ in the paramagnetic phase $g = 3.5$, close to the phase transition.
The isoTNS ansatz is optimized in several sweeps, each composed of one passing of the isometric center from the top left corner to the bottom right before returning it to the starting position.
We optimize each tensor in a column from top to bottom, then perform a Moses move to bring the isometric center to the top of the next column.

The construction of the quantum circuit representation of the isoTNS is detailed in the Supplementary Material.
We use a straightforward implementation that assigns one qubit to each physical output of the isoTNS, requiring a total of $N = (L_x+2) \times L_y$ qubits.
Transpilation on a quantum processor with a square grid connectivity results in a modest $200-300$ CNOT depth for a $5\times 5$ problem size, expected to scale like $\mathcal{O}(L_x+L_y)$.
More advanced strategies involving mid-circuit measurements and qubit reuse can be employed to lower the number of qubits from $\mathcal{O}(L_x \times L_y)$ to $\mathcal{O}(L_x+L_y)$ at the expense of an increased circuit depth \cite{Malz2025}.

\begin{figure}[t]
  \centering
  \includegraphics[width=0.48\textwidth]{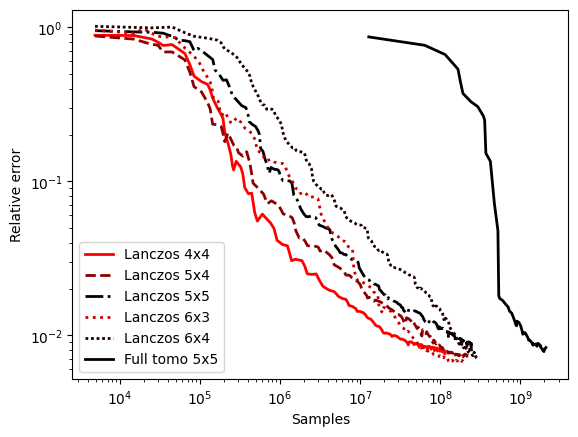}
  \caption{Ground-state energy optimization profile obtained for the $g=3.5$ two-dimensional transverse field Ising model.
  The quoted lattice sizes represent the physical lattice size (the isoTNS rows are 2 sites shorter).
  Results obtained with the Lanczos method for different lattice sizes are reported as a function of total sample cost.
  The profile obtained for the $5\times 5$ full tomography approach is also shown for comparison.}
  \label{fig:resultsLancz}
\end{figure}

We report the results obtained with the tomography method in Fig.~\ref{fig:resultsTomo}.
The various optimization profiles are obtained by sampling the tomography circuits using an exact, noiseless circuit simulator.
We compare these profiles to the optimization curve obtained by constructing the effective Hamiltonian with an exact tensor contraction.
The exact profile is almost perfectly reproduced using a number of shots per tomography operator of $\sim 10^5$.
The occasional increases in energy are due to the errors introduced when shifting the central column through the Moses move.

Figure ~\ref{fig:resultsLancz} illustrates the performance of the Lanczos method.
Compared to the tomography method, this optimization protocol is expected to converge in a larger number of sweeps.
This is because the eigenvalue problem is only solved approximately at each optimization step, requiring several passes through the same site to converge.
In order to mitigate the resulting shot overhead, we employ an adaptive sampling strategy: the number of measurements per circuit is doubled when an optimization step yields a worse energy than the previous one.
This approach avoids oversampling during early sweeps, further improving its shot cost efficiency.
As shown in Figure~\ref{fig:resultsLancz}, the Lanczos method reduces the number of shots until convergence by an order magnitude compared to the optimization curve of the $5\times 5$ tomography protocol.

Importantly, the number of samples required to achieve a given relative accuracy increases modestly with the system size.
This aligns with our theoretical prediction that the number of shots per tensor optimization depends only on the correlation length, not the full system size, and confirms the potential for scaling our proposed method to large system sizes.

\textit{Discussion}---We present an efficient algorithm for approximating the ground state of correlated 2D quantum materials, building on recent advances in tensor networks \cite{Zaletel2020} and quantum computing \cite{Schuhmacher2025}.
Compared to Monte-Carlo methods \cite{Foulkes2001}, our algorithm does not suffer from the sign problem.
Our approach encodes the ground-state wave function as a 2D isoTNS, and evaluates the classically-hard tensor contractions using quantum computers.
Moreover, it returns an efficient classical wave function representation alongside the energy.
Due to the isometric nature of the ansatz, the underlying classical description allows for easy preparation on a quantum register \cite{Slattery2021, Malz2025}.
Our numerical simulations show that the algorithm inherits the fast convergence and robustness of DMRG.
Once fully fault tolerant devices are available, this can serve as a good state preparation subroutine for quantum phase estimation \cite{Nelson2024}, if the exact ground state energy is required.

While we report results obtained with a noiseless simulator, the architecture, circuit depth, and shot requirements outlined are compatible with near-term and early fault-tolerant quantum hardware.
The largest bottleneck is the number of samples required to obtain an accurate energy estimate.
This sampling overhead can potentially be improved by reusing shots taken in previous sweeps, or by integrating information from approximate classical contractions to reduce estimator variance.

The circuit depth can be reduced using approximate compilation strategies \cite{Rakyta2022}.
This is of particular importance in the near future to mitigate gate errors present in real devices.
Nevertheless, we expect our algorithm to be relatively robust against noise.
In fact, local depolarizing noise will reduce the amplitude of all entries of the effective Hamiltonian by a similar factor.
Under this noise model, the correct optimal eigenvector is still recovered exactly---at the only price of an increased number of shots required for estimating the expectation values with a target accuracy.
In the presence of other error types (e.g., Sparse Pauli-Lindblad), more complex error mitigation schemes may be employed \cite{Cai2023, VanDenBerg2023_PEC,fischer2024dynamical}.

The algorithm can be generalized to three-dimensional systems and above, although the accuracy of the Moses move in this regime is largely unknown.
Our algorithm can, in principle, be extended to non-isometric PEPS, where sequential tensor updates are performed without any structural transformations of the underlying tensor network.
However, this requires preparing a generic PEPS on a quantum device.
This can be done using the Schwarz-Temme-Verstraete algorithm \cite{Schwarz2012}, albeit with much more complex circuits. 
Together with the effective Hamiltonian, this case also requires estimating an effective norm matrix that appears in the denominator of Eq.~\eqref{eq:energy}.
The resulting generalized eigenvalue problem may pose challenges if the norm matrix is ill-conditioned.

More generally, it is interesting to speculate whether the circuit optimization strategy used here could be applied to other classes of parameterized quantum circuits.
The reference optimization method -- global gradient descent through parameter space of a fixed ansatz \cite{Tilly2022} -- is hampered by the barren plateau problem \cite{McClean2018, Cerezo2021, Cunningham2025}.
Tensor network optimizers avoid this by solving a series of smaller eigenvalue problems, connected through shifts of the isometric center.
This strategy is known to be classically efficient for topologies without loops and -- as proven by our work -- in more general settings with access to a quantum computer.
Trying to understand if this paradigm, characterized by batch-wise parameter updates interspersed with topology transformations, can be applied elsewhere is a fascinating avenue for future work.

\textit{Acknowledgments}---This research was supported by NCCR MARVEL, a National Center of Competence in Research, funded by the Swiss National Science Foundation (grant number 205602) and by RESQUE funded by the Swiss National Science Foundation (grant number 225229).

\bibliographystyle{apsrev4-2}
\bibliography{bibliography}

\newpage

\title{Supplemental Material for "Optimizing two-dimensional isometric tensor networks with quantum computers"}

\author{Sebastian Leontica}

\date{\today}
\UseRawInputEncoding

\maketitle
\onecolumngrid

\section*{Circuit construction}


We describe here how to build a circuit that efficiently implements the isometry ($\mathcal{U}$ in the main text) mapping the legs of the isometric center to the physical qubits.

\begin{figure}[t]
  \centering
  \includegraphics[width=0.8\textwidth]{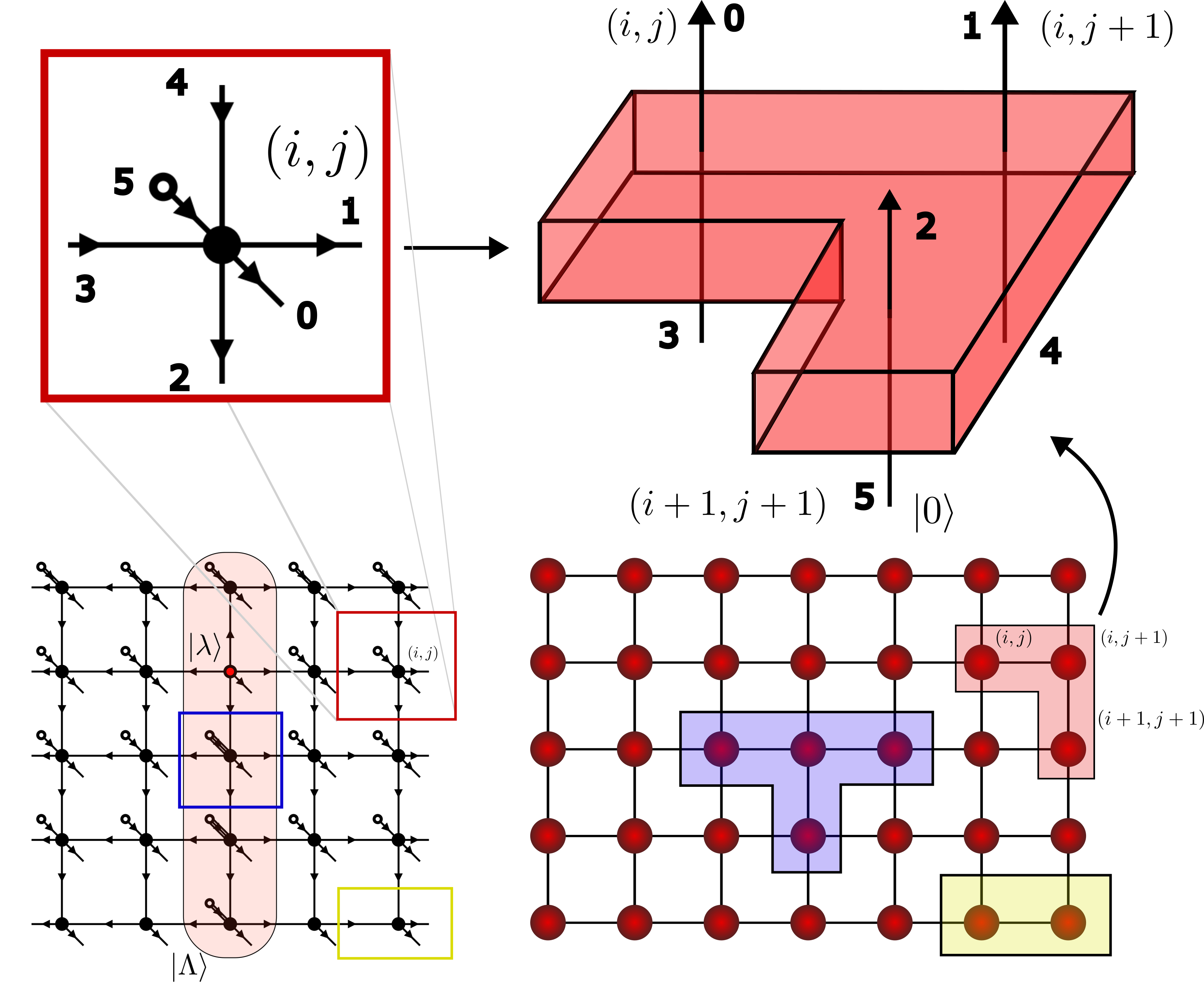}
  \caption{Representation of the circuit diagram associated to a node in the isoTNS.
  The red L-shaped block in the upper right corner represents a unitary transformation that embeds the isometry of the isoTNS associated with a site $(i, j)$ in the bulk of the physical lattice.
  Other unitary operations associated with nodes of the lattice are also marked with matching colors in the isoTNS diagram (bottom left) and the corresponding position in the physical lattice (bottom right).
  The blue tensor represents an isometry located along the central column.
  The yellow tensor represents an isometry located at the bottom edge.
  Note that the physical lattice is $7\times 5$, while the isoTNS is $5 \times 5$.
  This is because the tensors at the edge are associated with two physical indices (as discussed in the main text).}
  \label{fig:circuit}
\end{figure}

Consider a two-dimensional grid of qubits representing the physical degrees of freedom on which the Hamiltonian is defined.
Following the convention described in the main text, outside the central column, each isometry located in the lattice bulk have 3 input legs and 3 output legs.
Note that the value of one of the input legs is fixed to 0.
The action of the corresponding tensor in the circuit is illustrated in Fig.~\ref{fig:circuit}.
The qubits in the grid are labelled according to the site at which they are output in the isoTNS representation.
Note that the qubit grid includes two additional columns compared to the isoTNS.
As described in the main text, this is due to the fact that the isometries at the edge of the grid are associated to two physical indices.
Each isometry tensor can be embedded in a unitary operator and directly compiled in terms of elementary quantum gates.
The key challenge associated with mapping the overall isoTNS to a quantum circuit is to fix a convention for mapping the qubits to the input and output legs associated to a given unitary.
We adopt the mapping depicted in the upper part of Fig.~\ref{fig:circuit}.
This mapping is especially convenient because it translate each unitary translates into an L-shaped interaction between adjacent qubits when transpiled on a processor with a square connectivity.
To better understand the overall circuit structure, we trace the path of a qubit at $(i+1,j+1)$ in the physical lattice through the tensor diagram.
After being initialized to zero (index 5) by the tensor at position $(i,j)$ in the isoTNS, it is mapped to index $2$ ($5\to 2$).
Index $2$ is contracted with index $4$ of the tensor below (see left panel of Figure~\ref{fig:circuit}).
Hence, after that isometry is applied, the qubit is mapped to index $1$.
A third isometry then maps it as $3 \to 0$ -- \textit{i.e.} to the physical qubit associated with site $(i+1, j+1)$.
The same circuit diagram is mirrored to the left of the central column.
Qubits on the central column transverse only two isometries -- moving up or down one site through the tensor diagram depending on whether they are above or below the isometric center.
As an example, assume that the isometric center is located above row $i$ on column $j$.
A qubit is initialized at $(i,j)$ in the tensor network grid, travels down, and is returned at site $(i+1,j)$.
These rules allow for a complete assignment of qubit labels to the edges of the isoTNS.

All tensors in the bulk are isometries with $1+4\log_2 D$ indices (one physical and 4 auxiliary with dimension $\log_2 D$).
Their optimal compilation yields a $CX$ count per unitary of $\sim 2^{1+4\log_2 D} = 2D^4$ \cite{Iten2016}, without accounting for architectural constraints of the specific quantum platform.
The circuit representation of the full isoTNS contains $CX_{count} = \mathcal{O}(ND^4)$ entangling gates with a depth of $CX_{count} = \mathcal{O}((L_x+L_y) D^4)$.

\section*{Parameter shift rule}

Let $\mathcal{U}$ be the isometry mapping the state associated with the isometric center $\ket{v}$ to the physical output legs of the isoTNS, and $H$ the physical Hamiltonian.
We aim at designing a quantum protocol for estimating the result of applying the effective Hamiltonian, defined by $H_\text{eff} = \mathcal{U}^\dagger H \mathcal{U}$, to arbitrary states $\ket{v}$ as
\begin{equation}
    \ket{v'} = \mathcal{U}^\dagger H \mathcal{U} \ket{v}.
    \label{eq:Lanczos1}
\end{equation}
This serves as the primitive for the Lanczos method discussed in the main text.
Let $V$ be a unitary that prepares the current state of the isometric center on a quantum register such that $\ket{v} = V\ket{0}$.
In practice, among the possible unitary embeddings, we choose the shallowest circuit representation of $V$ without using ancillae or measurements \cite{Iten2016}.
For reasons that will become apparent soon, it is easier to compute the entries of an operator in the computational basis.
For this reason, instead of calculating the representation of $\mathcal{U}^\dagger H \mathcal{U}$ in the $\ket{v}$ basis, we revolve $V$ in the effective Hamiltonian.
In this way, the input state is $\ket{0}$ and the entries in the output state
\begin{equation}
    \ket{w} = V^\dagger \mathcal{U}^\dagger H \mathcal{U} V \ket{0},
    \label{eq:Lanczos2}
\end{equation}
are connected to those of the original vector $\ket{v'}$ (see Eq.~(\ref{eq:Lanczos1})) by a change of basis via $V$
\begin{equation}
    \ket{v'} = V\ket{w}.
\end{equation}
Note that the rotation is performed in the $2D^4$ dimensional space of the isometric center (in the bulk).
Hence, the above manipulation can be performed efficiently on a classical processor.
As we will show in the following, Eq.~(\ref{eq:Lanczos2}) is easier to evaluate than Eq.~(\ref{eq:Lanczos1}) on a quantum computer.

Let $X$ be the Pauli-X operator acting on one of the legs of the isometric center.
We consider the expectation value
\begin{equation}
    E(\theta) = \bra{0}e^{-i\frac{\theta}{2} X}V^\dagger \mathcal{U}^\dagger H \mathcal{U} V e^{i\frac{\theta}{2} X} \ket{0}.
    \label{eq:Lanczos_ExpVal}
\end{equation}

The parameter shift rule \cite{Wierichs2022} yields
\begin{equation}
 \begin{aligned}
    \frac{1}{2}(E(\frac{\pi}{2})- E(-\frac{\pi}{2})) 
      &= -\frac{i}{2}(\bra{0}XV^\dagger \mathcal{U}^\dagger H \mathcal{U} V\ket{0}
                    -\bra{0}V^\dagger \mathcal{U}^\dagger H \mathcal{U} VX\ket{0}) \\
      &= \Im (\bra{0}XV^\dagger H_\text{eff}\ket{v}) \\
      &= \Im (\bra{0} X \ket{w}) \, .
 \end{aligned}
\end{equation}

Hence, the imaginary part of the projection of $\ket{w}$ along the basis vector $X\ket{0}$ can be calculated from $E(\theta)$ evaluated at $\theta = \pm \frac{\pi}{2}$.
We obtain the projection along the remaining (rotated) basis vectors by preparing the input states
\begin{align}
    \ket{\phi^R_{\pm,s}} &= \frac{1}{\sqrt{2}}(\ket{0}\pm\ket{s}), \\
    \ket{\phi^I_{\pm,s}} &= \frac{1}{\sqrt{2}}(\ket{0}\pm i\ket{s}),
\end{align}
where $s$ is an arbitrary binary string over the legs of the isometric center.
These states can be prepared efficiently using standard GHZ circuits.
All coefficients in the expansion of $\ket{w}$ in the computational basis can be measured using the parameter shift rule above, as follows:
\begin{equation}
\label{eq:coeffs_Lanczos}
\begin{split}
    \Re(\bra{s}V^\dagger H_\text{eff}\ket{v}) &= \Re(\bra{s} \ket{w})
      = \frac{1}{2}(E^R_{+,s}-E^R_{-,s}), \\
    \Im(\bra{s}V^\dagger H_\text{eff}\ket{v}) &= \Im(\bra{s} \ket{w})
      = \frac{1}{2}(E^I_{+,s}-E^I_{-,s}),
\end{split}
\end{equation}
where all terms on the RHS are expectation values of the rotated effective Hamiltonian $V^\dagger \mathcal{U}^\dagger H\mathcal{U} V$ in the corresponding input state.

\section*{Error analysis}

We analyse the impact of shot-noise error on the accuracy of the ground-state energy estimate of the effective Hamiltonian, using both tomography and the Lanczos approach.
We consider Hamiltonians composed of terms geometrically localized around some position $\vec{r}$ in the lattice (a condition satisfied by many lattice Hamiltonians).
Concretely, we work with Hamiltonians of the form
\begin{equation}
\label{eq:physHamilt}
    H = \sum_{\alpha, \Vec{r}} h_\alpha O_\alpha^{(\Vec{r})},
\end{equation}
where $\alpha$ is an index for the classes of mutually commuting Pauli terms that can be measured simultaneously on the device, and $O_{\alpha}^{(\Vec{r})}$ is a Pauli operator of the $\alpha$-th group centered at position $\Vec{r}$ in the lattice. 

\subsection*{Hamiltonian tomography}

\begin{figure}[t]
  \centering
  \includegraphics[width=0.6\textwidth]{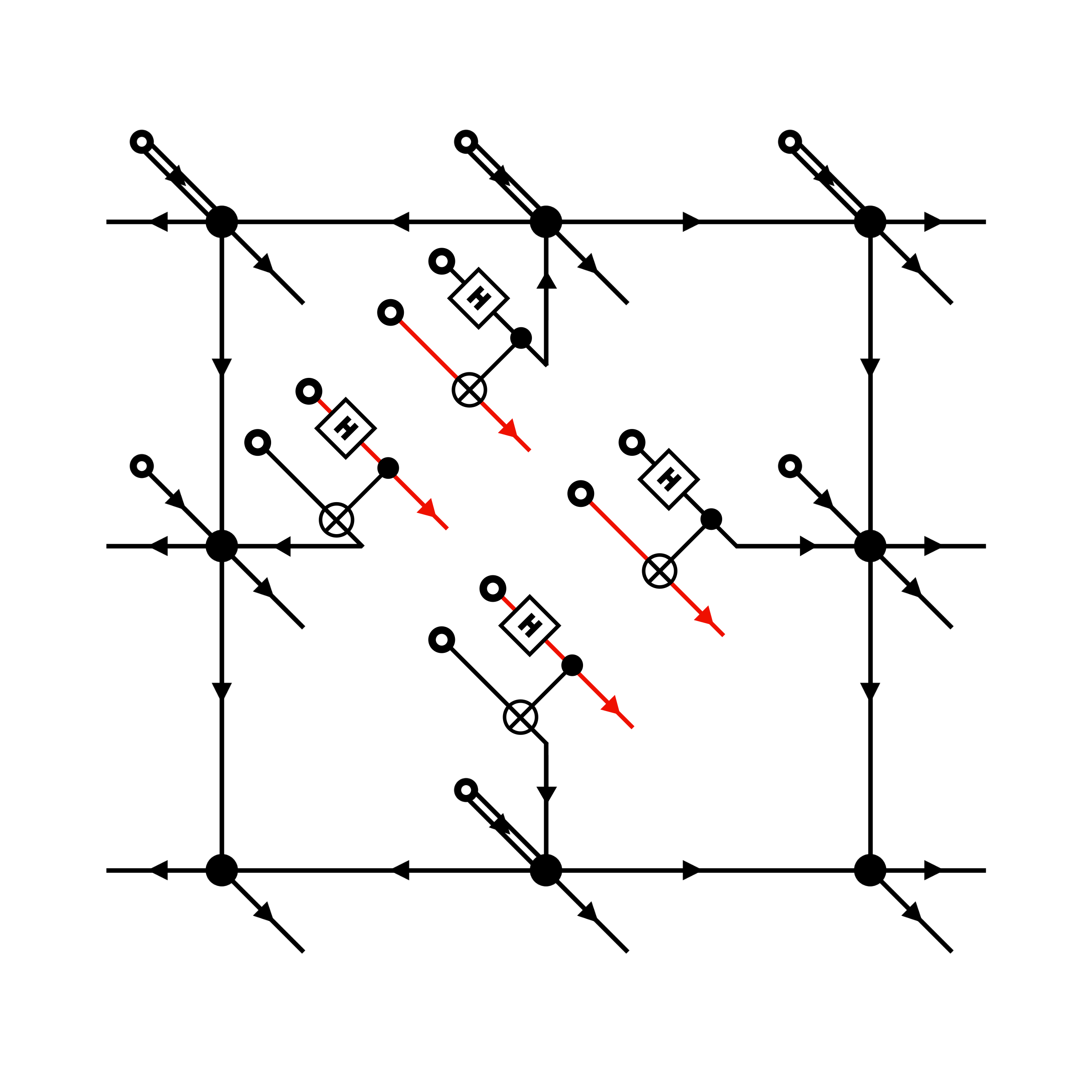}
  \caption{Illustration of the circuits used in the tomography method for a $3\times 3$ isoTNS. The output is the state $\ket{\tilde{\Psi}}$ defined in the main text. The auxiliary space on which tomography is performed is marked with red qubit output lines. The remaining outputs of the initial Bell pair production step are the input of the isoTNS isometry $\mathcal{U}$ and subsequently mapped to a state in physical space (black output lines). The physical leg of the isometric center itself is decoupled from the other legs, so it is omitted here.}
  \label{fig:bellpairs}
\end{figure}

Since the Pauli operators form a complete basis in operator space, it is possible to decompose the effective Hamiltonian as
\begin{equation}
    H_\text{eff} = \sum_P c_P P,
    \label{eq:HEff_Pauli_Expansion}
\end{equation}
where $c_P$ are real coefficients and $P$ labels the Pauli operators acting on the same space on which the isometric center $\ket{\lambda}$ is defined.
If $Q$ is an arbitrary Pauli operator taken from this set, we have
\begin{equation}
    \Tr\left(QH_\text{eff}\right) = \sum_P c_P \Tr(QP) = \sum_P c_P 2D^4 \delta_{PQ} = 2D^4 c_Q,
\end{equation}
where $\Tr I = 2D^4$ is the dimension of the space (for an isometric center in the bulk of the isoTNS).
This shows that the coefficients $c_P$ in Eq.~(\ref{eq:HEff_Pauli_Expansion}) are
\begin{equation}
    H_\text{eff} = \frac{1}{2D^4}\sum_P \Tr(P H_\text{eff}) P.
\end{equation}
In the tomography protocol, one needs to evaluate $\Tr(P H_\text{eff})$ on the quantum processor for each Pauli operator $P$.
To this end, we introduce the maximally-entangled state,
\begin{equation}
    \ket{\Phi} 
        = \frac{1}{\sqrt{2D^4}} \sum_{i=0}^{2D^4-1} \ket{i}\otimes \ket{i} \, .
\end{equation}
Standard quantum information identities allow us to rewrite the trace as
\begin{equation}
    \frac{1}{2D^4}\Tr(PH_\text{eff}) = \bra{\Phi} I\otimes (PH_\text{eff})\ket{\Phi} = \bra{\Phi} P^T \otimes H_\text{eff}\ket{\Phi}.
\end{equation}
Reintroducing the definition of the effective Hamiltonian in terms of the physical Hamiltonian $H_\text{eff} = \mathcal{U}^\dagger H\mathcal{U}$ we arrive at the equation for the coefficients reported in the main text
\begin{equation}
  c_P = \bra{\Phi} P^T \otimes H_\text{eff}\ket{\Phi} 
      = \bra{\tilde{\Psi}}P^T \otimes H \ket{\tilde{\Psi}}.
  \label{eq:CoefficientExpansion}
\end{equation}
where $\ket{\tilde{\Psi}}$ is defined in Eq.~(\ref{eq:Psi}) of the main text.
Substituting in Eq.~(\ref{eq:CoefficientExpansion}) the expression for the Hamiltonian given in Eq.~\eqref{eq:physHamilt} we derive the following expression for $H_\text{eff}$
\begin{equation}
    H_\text{eff} 
        = \sum_{P \neq I} \langle P^T \otimes H_\text{eff} \rangle_{\Phi} P 
        = \sum_{P \neq I} \langle P^T \otimes H \rangle_{\Tilde\Psi} P 
        = \sum_{P\neq I,\alpha} h_\alpha\left\langle P^T \otimes \left(\sum_{\Vec{r}} O_\alpha^{(\Vec{r})}\right)\right\rangle_{\Tilde{\Psi}} P 
        = \sum_{P \neq I} h_P P,
\end{equation}
where the constant shift term is dropped because it does not modify the optimal eigenvector.
We prepare the state $\ket{\Tilde{\Psi}}$ on the quantum computer (see Fig.~\ref{fig:bellpairs}) and measure $n_\alpha$ times all qubits in a basis that simultaneously diagonalizes all $P^T \otimes O_\alpha^{\Vec{r}}$ operators.
Since we assumed a Pauli decomposition, each term in the sum is $\pm 1$ for each shot.
The variance of the sum for a single shot measured for the $P$ operator and for the $\alpha$-th group of commuting term of the original Hamiltonian $H$ can be computed as
\begin{equation}
    \operatorname{Var}_{\Tilde{\Psi}}\left[\sum_{\Vec{r}} P^T \otimes O_\alpha^{(\Vec{r})}\right] = \sum_{\Vec{r},\Vec{r}'}\left( \langle O_\alpha^{\Vec{r}}O_\alpha^{\Vec{r}'} \rangle_{\Tilde{\Psi}} - \langle P^T \otimes O_\alpha^{\Vec{r}} \rangle_{\Tilde{\Psi}}\langle P^T \otimes O_\alpha^{\Vec{r}'} \rangle_{\Tilde{\Psi}}\right).
    \label{eq:VarianceNaive}
\end{equation}

We are primarily interested in how the variance scales with the system size.
The expression above includes $\mathcal{O}(N^2)$ terms, where $N = (L_x+2)L_y$ is the number of sites in the physical system.
If the system has a finite correlation length, the majority of the terms in the sum will be such that all the distances involved $\abs{\vec{r}}, \abs{\vec{r}'},\abs{\Vec{r}-\Vec{r}'}$ are much larger than the correlation length (where $\vec{r}, \vec{r}'$ are conventionally defined with the origin at the isometric center).
Moreover, $P$ is located at the isometry center (hence, the origin).
In the following, we will investigate if we can use the statistical independence of far-away operators to reduce this sum to only $\mathcal{O}(N)$ contributing terms.
In the aforementioned limit, the contributions of a given term can approximated as
\begin{equation}
    \langle O_\alpha^{\Vec{r}}O_\alpha^{\Vec{r}'} \rangle_{\Tilde{\Psi}} - \langle P^T \otimes O_\alpha^{\Vec{r}} \rangle_{\Tilde{\Psi}}\langle P^T \otimes O_\alpha^{\Vec{r}'} \rangle_{\Tilde{\Psi}} \approx 
    \langle O_\alpha^{\Vec{r}} \rangle_{\Tilde{\Psi}}
    \langle O_\alpha^{\Vec{r}'} \rangle_{\Tilde{\Psi}}
  - \langle P^T \rangle_{\Tilde{\Psi}} 
    \langle O_\alpha^{\Vec{r}} \rangle_{\Tilde{\Psi}}
    \langle P^T \rangle_{\Tilde{\Psi}} 
    \langle O_\alpha^{\Vec{r}'} \rangle_{\Tilde{\Psi}} = 
    \langle O_\alpha^{\Vec{r}}\rangle_{\Tilde{\Psi}} \langle O_\alpha^{\Vec{r}'} \rangle_{\Tilde{\Psi}},
    \label{eq:Variance_SingleTerm}
\end{equation}
as $\langle P^T\rangle_{\Tilde{\Psi}} \propto \Tr P = 0$ for $P \neq I$ by construction.
Since this term does not automatically vanish, all $\mathcal{O}(N^2)$ contributions to Eq.~(\ref{eq:VarianceNaive}) are potentially non-negligible.
This scaling can be reduced by working instead with centered operators $\Tilde{O}_\alpha^{(\vec{r})}$ defined as
\begin{equation}
    \Tilde{O}_\alpha^{(\vec{r})} = O_\alpha^{\vec{r}} 
        - \langle O_\alpha^{\vec{r}} \rangle_{\Tilde{\Psi}},
\end{equation}
where, in practice, the shift $\langle O_\alpha^{\vec{r}} \rangle_{\Tilde{\Psi}}$ is constructed through quantum measurements on a circuit preparing $\Tilde{\Psi}$ and is therefore potentially subject to shot noise.
We define as $\epsilon_\alpha^{\vec{r}}$ the difference between this estimate and its exact (noiseless) counterpart.
This leads to a new Hamiltonian that only differs from the original by a constant energy shift
\begin{equation}
    \tilde{H} = \sum_{\alpha,\vec{r}} h_\alpha \Tilde{O}_\alpha^{(\vec{r})} 
      = H - \sum_{\alpha,\vec{r}}\langle O_\alpha^{\Vec{r}}\rangle_{\Tilde{\Psi}},
\end{equation}
which does not affect its ground state and therefore recovers the same optimal isometric center tensor.

Since $\langle P^T\rangle_{\Tilde{\Psi}} = 0$, the Hamiltonian does not change when expressed in terms of the centered observables.
Hence, the estimator variance becomes
\begin{equation}
\label{eq:Var}
    \operatorname{Var}_{\Tilde{\Psi}}\left[\sum_{\Vec{r}} P^T \otimes \tilde{O}_\alpha^{(\Vec{r})}\right] 
        = \sum_{\Vec{r},\Vec{r}'}\left( \langle \tilde{O}_\alpha^{\Vec{r}}\tilde{O}_\alpha^{\Vec{r}'} \rangle_{\Tilde{\Psi}} - \langle P^T \otimes \tilde{O}_\alpha^{\Vec{r}} \rangle_{\Tilde{\Psi}}\langle P^T \otimes \tilde{O}_\alpha^{\Vec{r}'} \rangle_{\Tilde{\Psi}}\right) = \sum_{\Vec{r},\Vec{r}'}\langle \tilde{O}_\alpha^{\Vec{r}}\rangle_{\Tilde{\Psi}} \langle \tilde{O}_\alpha^{\Vec{r}'} \rangle_{\Tilde{\Psi}} + \mathcal{O}(N),
\end{equation}
where we used the fact that for Pauli operators $P^2 = I$, and the last equality holds when the correlation length is finite.
In practice, the derivation reported above indicates that the scaling of the variance in the asymptotic limit $\mathcal{O}(N^2)$ is governed by contributions from ``disconnected terms''.
These terms can be canceled out by appropriately shifting the operator.
The remaining correction terms appear only when at least two of the 3 operators involved ($O^{\vec{r}}_\alpha$, $O^{\vec{r}'}_\alpha$, $P^T$) are close together.
The number of these terms scales only like $\mathcal{O}(N)$.
Although $P$ is defined in the auxilliary space, its action is still localized around the position of the isometric center.
Therefore we expect $P$ to behave similarly to an operator in physical space located on the isometric center (conventionally $\vec{r} = 0$).
If we had perfect knowledge of the shifts $\langle O_\alpha^{\vec{r}} \rangle_{\Tilde{\Psi}}$, the first term in the last equality would vanish.
However, these quantities also need to be estimated from samples.
Using the previously defined $\epsilon_\alpha^{\vec{r}}$, the first term is
\begin{equation}
\label{eq:error_sum}
    \sum_{\Vec{r},\Vec{r}'}\langle \tilde{O}_\alpha^{\Vec{r}}\rangle_{\Tilde{\Psi}} \langle \tilde{O}_\alpha^{\Vec{r}'} \rangle_{\Tilde{\Psi}} = \left(\sum_{\vec{r}} \epsilon_{\alpha}^{\vec{r}}\right)^2.
\end{equation}

If we take $m_\alpha$ samples from $\tilde{\Psi}$ in the mutual basis of the $O_\alpha^{\vec{r}}$ operators, then the expectation value of the quantity in Eq.~\eqref{eq:error_sum} is
\begin{equation}
    \mathbb{E}\left[\sum_{\vec{r},\vec{r}'} \epsilon_\alpha^{(\vec{r})}\epsilon_\alpha^{(\vec{r}')}\right] = \frac{1}{m_\alpha} \sum_{\vec{r},\vec{r}'} \operatorname{Cov}_{\tilde{\Psi}}\left[O^{(\vec{r})}_\alpha, O^{(\vec{r}')}_\alpha \right] \sim \mathcal{O}\left(\frac{N}{m_\alpha}\right),
\end{equation}
where the asymptotic scaling assumes again that the system has finite correlation length, such that only terms with $\vec{r}$ and $\vec{r}'$ close together contribute. When $m_\alpha$ is large, the joint distribution of the errors $\epsilon_\alpha^{\vec{r}}$ for different $\vec{r}$ is Gaussian by the central limit theorem, such that their sum also follows a Gaussian distribution. The calculation above shows that its standard deviation scales like $\mathcal{O}(\sqrt{N})$. Large deviations of Gaussian variables are heavily suppressed, so the probability of an error on the scale of $\mathcal{O}(N)$ is exponentially small in $N$. Typical values of the error are then also expected to obey
\begin{equation}
    \left(\sum_{\vec{r}} \epsilon_{\alpha}^{\vec{r}}\right)^2 \sim \mathcal{O}\left(\frac{N}{m_\alpha}\right).
\end{equation}

Substituting this into Eq.~\eqref{eq:Var} we see that
\begin{equation}
    \operatorname{Var}_{\Tilde{\Psi}}\left[\sum_{\Vec{r}} P^T \otimes \tilde{O}_\alpha^{(\Vec{r})}\right] \sim \mathcal{O}(N),
\end{equation}
almost surely even with $\langle O_\alpha^{\vec{r}} \rangle_{\Tilde{\Psi}}$ estimated from samples. 
In practice, to avoid taking extra samples, we combine all tomography shots to obtain an accurate estimate of the shifts $\langle O_\alpha^{\vec{r}} \rangle_{\Tilde{\Psi}}$.
This is possible because the operator $O_\alpha^{\vec{r}}$ acts non-trivially only on the physical qubits, and trivially on the ancilla qubits on which tomography is performed.
The number of shots is then on the order of $m_\alpha = 4D^8 n_\alpha$, yielding a very accurate approximation of the shifts.
If $\xi$ is the correlation length of the system and $\eta$ is the critical exponent describing the decay of most 2-point correlators, we can estimate their contribution to the sum in Eq.~\eqref{eq:VarianceNaive} using the standard scaling law as
\begin{equation}
    \sum_{\vec{r},\vec{r}'} \operatorname{Cov}_{\tilde{\Psi}}\left[O_\alpha^{\vec{r}}, O_\alpha^{\vec{r}'}\right] \sim N\xi^{2-\eta}.
\end{equation}

We then expect the variance of our estimator for the coefficient $h_P$ in the expansion of $H_\text{eff}$ to be given by
\begin{equation}
\label{eq:var_hP}
    \operatorname{Var}_{\tilde{\Psi}}[\hat{h}_P] \sim c\frac{h^2 \xi^{2-\eta} N}{n},
\end{equation}
in the large $N$ limit, where $h$ is an energy scale of terms in the Hamiltonian and $n_{\alpha} = n$ is a common number of shots taken for each commuting group.

The result of the sampling procedure is an estimator $\hat{H}_\text{eff}$ of the true effective Hamiltonian $H_{\text{eff}}$.
Diagonalizing this classically returns an estimator of the optimal isometric center.
We quantify the shot noise error in the resulting energy estimate as
\begin{equation}
    \Delta = \bra{\hat{\lambda}_\text{min}}H_\text{eff}\ket{\hat{\lambda}_\text{min}} - \lambda_\text{min},
\end{equation}
where we denote by $\ket{\hat{\lambda}_{\min}}$ the ground state eigenvector of the estimated Hamiltonian $\hat{H}_\text{eff}$, and by $\lambda_\text{min}$ the exact ground state eigenvalue of $H_\text{eff}$.
Note that while $\ket{\hat{\lambda}_{\min}}$ is known during the optimization, this is not the case for $H_\text{eff}$ and $\lambda_\text{min}$.
Hence, the actual error cannot be calculated efficiently in practice.
When the ansatz is relatively well converged, the gap of $H_\text{eff}$ should be at least as large as the real gap $\delta$ of the model.
This can be justified as follows: if the lowest-lying eigenvector $\ket{\lambda_{\min}}$ is mapped to the true ground state under the rest of the isoTNS $\ket{\Psi} = \mathcal{U}\ket{\lambda_{\min}}$, then the first excited state $\ket{\lambda_1}$ is mapped to a vector in physical space that is orthogonal to the true ground state.
Since this vector has no support on the ground state, the energy of $\mathcal{U}\ket{\lambda_1}$ must be at least as large as the energy of the first excited state of $H$.
We denote the error in the effective Hamiltonian estimator by $\delta H = \hat{H}_\text{eff} - H_\text{eff}$ and let $\ket{\phi}$ be a quantum state on the support of $H_\text{eff}$.
Under the assumption that the system is not critical and has a finite gap $\delta$, we can then derive the following bound on $\Delta$ using the Temple inequality
\begin{equation}
\label{eq:errtomo}
    \Delta \leq \frac{\norm{\left(H_\text{eff}-\bra{\hat{\lambda}_\text{min}}H_\text{eff}\ket{\hat{\lambda}_\text{min}} \mathbb{I} \right)\ket{\hat{\lambda}_\text{min}}}^2}{\delta-\Delta} \approx \frac{\max_{\phi}\left(\langle\delta H^2\rangle_\phi-\langle\delta H\rangle_\phi^2\right)}{\delta}.
\end{equation}

We assume that sufficiently many shots are taken so that $\Delta \ll \delta$, i.e. the error in energy due to sampling noise is much less than the gap.
The numerator can be computed directly in terms of the spectrum of $\delta H$
\begin{equation}
    \max_{\phi}\left(\langle\delta H^2\rangle_\phi-\langle\delta H\rangle_\phi^2\right) = \frac{(\theta_{\max}-\theta_{\min})^2}{4},
\end{equation}
where $\theta_{\max}$ and $\theta_{\min}$ are the largest and smallest eigenvalues of $\delta H$, respectively.
The coefficients of $\delta H$ in the Pauli basis are generated solely by shot noise error, and therefore given by independent Gaussian random variables of mean $0$ and standard deviation $\sigma$ that can be approximated from Eq.~\eqref{eq:var_hP} as
\begin{equation}
    \sigma \approx \sqrt{ \frac{h^2 \xi^{2-\eta}N}{n}},
\end{equation}
possibly up to a numerical prefactor.
The statistics of eigenvalues for such an isotropic Gaussian matrix is given by the semicircle law, with a radius $R = 2\sigma\times (2D^4)$. We can then further approximate
\begin{equation}
    \frac{(\theta_{\max}-\theta_{\min})^2}{4} \approx R^2 = 16\sigma^2 D^8,
\end{equation}
leading to the following bound on the error
\begin{equation}
    \Delta \leq 16D^8\frac{h^2\xi^{2-\eta}N}{n \delta}.
\end{equation}

Hence, to obtain a fixed relative error $\epsilon_r \approx \Delta/(h^2N)$ in the ground-state energy we require a number of shots per tomography operator scaling like
\begin{equation}
    n = \mathcal{O}\left(\frac{D^8\xi^{2-\eta}}{\epsilon_r}\right).
\end{equation}

The full tomography requires $\mathcal{O}(D^8)$ operators.
Therefore, the total number of shots needed for one optimization step is given by
\begin{equation}
    n_{tot} \approx D^8 n = \mathcal{O}\left( \frac{D^{16} \xi^{2-\eta}}{\epsilon_r}\right).
\end{equation}

\subsection*{Lanczos method}

The variance of each estimator $\hat{E}$ in Eq.~\eqref{eq:coeffs_Lanczos} corresponding to different strings $s$ used when constructing $\ket{w}$ is given by
\begin{equation}
    \operatorname{Var}_{\phi}\left[\hat{E}\right] = \frac{1}{n} \sum_{\alpha,\Vec{r}, \Vec{r}'} h_\alpha^2\operatorname{Cov}_\phi\left[O_\alpha^{(\Vec{r})},O_\alpha^{(\Vec{r}')}\right] \sim h^2\xi^{2-\eta} \frac{N}{n},
    \label{eq:errLancz_0}
\end{equation}
where $n$ is the number of samples taken per expectation---value estimation in the protocol, and the final linear asymptotic scaling with system size holds for systems away from criticality with coherence length $\xi$. Then our estimator for $\ket{w} = V^\dagger H_\text{eff}\ket{v}$ is
\begin{equation}
    \ket{\hat{w}} = \ket{w} + \delta \ket{w},
\end{equation}
where $\delta \ket{w}$ is a complex random vector with independent Gaussian distributed entries of mean $0$ and variance given by the expression above.
Note that the first entry $\bra{0}\ket{w} = \bra{0}V^\dagger H_\text{eff} V\ket{0}$ is real and can be estimated directly, while the remaining entries are complex-valued and estimated based on 4 separate estimators.
In a crude approximation, performing the Lanczos approach based on the noisy results is equivalent to finding the exact ground state of a Hamiltonian with noisy entries.
Using the Temple inequality again in a similar fashion as above we get the following error estimate
\begin{equation}
\label{eq:errLancz}
    \Delta \sim \frac{D^4 h^2\xi^{2-\eta}N}{n \delta}.
\end{equation}

The difference in scaling between Eq.~\eqref{eq:errLancz} and Eq.~\eqref{eq:errtomo} ($D^4$ rather than $D^8$) appears because the $\ket{w}$ state has $D^4$ entries, each one affected by the error given in Eq.~(\ref{eq:errLancz_0}).

For a fixed relative error $\epsilon_r \approx \Delta/(h^2N)$ using this approach we require a total number of shots that scales like
\begin{equation}
    n_{tot} \approx D^4 n = \mathcal{O}\left(\frac{D^{8}}{\epsilon_r}\right).
\end{equation}

It is worth noting that both the Hamiltonian tomography and the Lanczos methods allow full parallelization of the shot collection, as all the expectation values required are independent and can be computed simultaneously.
This can help mitigate the large bond dimension scaling of this part of the algorithm.

\end{document}